\documentclass[acmtog]{acmart}

\AtBeginDocument{%
  }

\acmDOI{}

\acmConference[]{}{}{}
\acmPrice{}
\acmISBN{}

\usepackage[ruled,vlined,linesnumbered]{algorithm2e}
\SetCommentSty{small}
\usepackage{tikz}
\usepackage{pgf}
\usepackage{multirow,tabularx}
\usepackage{booktabs}
\usepackage{subfigure}
\usepackage{xcolor,colortbl}
\definecolor{lavender}{rgb}{0.9, 0.9, 0.98}
\definecolor{asparagus}{rgb}{0.53, 0.66, 0.42}
\definecolor{caribbeangreen}{rgb}{0.0, 0.8, 0.6}
\definecolor{darkolivegreen}{rgb}{0.33, 0.42, 0.18}
\definecolor{darkpastelgreen}{rgb}{0.01, 0.75, 0.24}
\begin{document}
\title{One-Shot Traffic Assignment with Forward-Looking Penalization}

\author{Giuliano Cornacchia}
\affiliation{
\institution{University of Pisa and ISTI-CNR}
\city{Pisa}
\country{Italy}
}
\email{giuliano.cornacchia@phd.unipi.it}

\author{Mirco Nanni}
\affiliation{
   \institution{ISTI-CNR}
\city{Pisa}
\country{Italy}} \email{mirco.nanni@isti.cnr.it}

\author{Luca Pappalardo}
\affiliation{
\institution{ISTI-CNR}
\city{Pisa}
\country{Italy}
}
\email{luca.pappalardo@isti.cnr.it}

\renewcommand{\shortauthors}{Cornacchia et al.}

\begin{abstract}
Traffic assignment (TA) is crucial in optimizing transportation systems and consists in efficiently assigning routes to a collection of trips. 
Existing TA algorithms often do not adequately consider real-time traffic conditions, resulting in inefficient route assignments. 
This paper introduces METIS, a cooperative, one-shot TA algorithm that combines alternative routing with edge penalization and informed route scoring. 
We conduct experiments in several cities to evaluate the performance of METIS against state-of-the-art one-shot methods. 
Compared to the best baseline, METIS significantly reduces CO2 emissions by 18\% in Milan, 28\% in Florence, and 46\% in Rome, improving trip distribution considerably while still having low computational time.
Our study proposes METIS as a promising solution for optimizing TA and urban transportation systems.
\end{abstract}





\maketitle

\section{Introduction}
\label{sec:introduction}
Traffic Assignment (TA) has emerged as a crucial problem today due to the rapid growth of urbanization and increasing traffic congestion \cite{campbell1950route, wang2018dynamic, chen1991ita, wardrop1952some, lujak2015route, beckman1956studies}. 
As cities expand and populations rise, transportation networks face pressure to efficiently accommodate the growing demand for mobility. 
Efficient TA plays a pivotal role in achieving several Sustainable Development Goals (SDGs) set by the United Nations \cite{assembly2015sustainable}, promoting effective traffic management and reducing greenhouse gas emissions.

Existing approaches to TA can be broadly classified into one-shot and iterative methods.
One-shot approaches assign routes to a collection of trips without any additional optimization \cite{campbell1950route, chen1991ita}, while iterative approaches involve multiple iterations to improve efficacy \cite{wardrop1952some, lujak2015route, beckman1956studies}.
However, these approaches predominantly rely on basic road network information and travel times, failing to harness the potential of more sophisticated measures based on mobility patterns. 
As a result, there is ample opportunity for further advancements of TA solutions to enhance their effectiveness.

In contrast to one-shot and iterative methods, alternative routing (AR) methods adopt an individualistic approach. 
They focus on providing alternative routes to individual users, aiming to strike a balance between proximity to the fastest path and route diversity \cite{li2022comparing, yen1971finding, aljazzar2011k, cheng2019shortest, suurballe1974disjoint, camvit2005choice, liu2018finding, chondrogiannis2015kshortest, chondrogiannis2018mincoll, hacker2021most}. 
However, their individualistic nature overlooks vehicle interactions, leading to suboptimal outcomes at the collective level. 
As a result, they often lead to increased congestion and a higher environmental impact. 

To overcome these limitations, we propose METIS, a novel cooperative approach that improves TA by incorporating alternative routing, edge penalization, and informed route scoring.
METIS introduces some key innovations. 
Firstly, METIS estimates vehicles' current position to penalize road edges expected to be traversed, discouraging future vehicles from using those congested edges.
Secondly, METIS generates alternative routes using the penalized road network and assigns them to individual trips favouring unpopular routes with high-capacity roads. 
These innovative components enable METIS to promote a more balanced distribution of traffic, improving the efficiency of TA and providing drivers with fast paths while addressing the limitations of existing approaches.

Through comprehensive experiments conducted in three cities, we provide compelling evidence of METIS's effectiveness in reducing the environmental impact of traffic, particularly CO2 emissions. 
By comparing METIS with various state-of-the-art approaches, including individualistic and collective one-shot methods, we highlight its superior performance in optimizing routing while maintaining computational efficiency.
Notably, METIS significantly reduces total CO2 emissions compared to the best baseline, ranging from 18\% to 46\%, depending on the city.

METIS represents a significant step forward in TA, offering a cooperative and dynamic approach to guide drivers towards efficient routes and alleviate congestion in urban areas.
The key contributions of this paper can be summarized as follows:

\begin{itemize}
\item We introduce Forward-Looking Edge Penalization (FLEP) to estimate vehicles' current positions and penalize road edges that are expected to be traversed (Section \ref{sec:flep});
\item We integrate AR into TA, showing how generating alternative routes may improve traffic assignment (Section \ref{sec:kmdnsp});
\item We introduce a pattern-based route scoring to discourage the selection of popular, congested routes (Section \ref{sec:route_gen}).
\item We conduct extensive experiments and simulations, comparing AR solutions, one-shot approaches, and our METIS algorithm in three cities, demonstrating the superior performance of METIS in reducing CO2 emissions while maintaining competitive computational performance (Section \ref{sec:results}).
\end{itemize}

\subsubsection*{\bf Open Source}
The code that implements METIS, the baselines, and the experiments can be accessed at \url{https://bit.ly/metis_ta}. 

\section{Related Work}
\label{sec:related}
Traffic assignment (TA) consists in allocating vehicle trips on a road network to minimize congestion and travel time \cite{campbell1950route, wang2018dynamic, chen1991ita, wardrop1952some, lujak2015route, beckman1956studies}. 
We group TA solutions into individual approaches, providing a route to a single trip, and collective approaches, providing a set of routes for an entire collection of trips.

\subsection*{Individual approaches}
The fastest path is the most straightforward approach to connect two locations in a road network \cite{wu2012shortest}. 
However, from a collective point of view, aggregating all individual fastest paths may increase congestion and CO2 emissions \cite{cornacchia2022how}.

Several works focus on alternative routing (AR) to distribute the vehicles more evenly on the road network \cite{li2022comparing}. 
In particular, the $k$-shortest path problem \cite{yen1971finding, aljazzar2011k} aims to find the $k$ shortest paths between an origin and a destination.
In practical scenarios, $k$-shortest path solutions fail to provide significant path diversification, as the generated paths exhibit a 99\% overlap in terms of road edges \cite{cheng2019shortest}.
The $k$-shortest disjointed paths problem \cite{suurballe1974disjoint} focuses on identifying $k$ paths that do not overlap. 
Solutions to this problem often result in routes that significantly deviate from the optimal path, leading to a notable increase in travel time. 
Several approaches lie between the $k$-shortest path and $k$-shortest disjoint paths, which can be divided into edge weight, plateau, and dissimilarity approaches.

\paragraph{Edge weight approaches} They compute the shortest paths iteratively. 
At each iteration, they update the edge weights of the road network to compute $k$ alternative paths. 
Edge weight updating may consist of a randomization of the weights or a cumulative penalization of the edges contributing to the shortest paths. 
Although easy-to-implement, edge-weight approaches do not guarantee the generation of paths considerably different from each other \cite{li2022comparing}. 

\paragraph{Plateau approaches.} 
They build two shortest-path trees, one from the source and one from the destination, and identify their common branches, known as plateaus \cite{camvit2005choice}. 
The top-$k$ plateaus are selected based on their lengths, and alternative paths are generated by appending the shortest paths from the source to the first edge of the plateau and from the last edge to the target. As the plateaus are inherently disjointed, they may create significantly longer routes than the fastest path \cite{camvit2005choice}.

\paragraph{Dissimilarity approaches.} They generate $k$ paths that satisfy a dissimilarity constraint and a desired property. 
Liu et al. \cite{liu2018finding} propose the $k$-Shortest Paths with Diversity ($k$SPD) problem, defined as top-$k$ shortest paths that are the most dissimilar with each other and minimize the paths' total length. 
Chondrogiannis et al. \cite{chondrogiannis2015kshortest} propose an implementation of the $k$-Shortest Paths with Limited Overlap ($k$SPLO), seeking to recommend $k$-alternative paths that are as short as possible and sufficiently dissimilar. 
Chondrogiannis et al. \cite{chondrogiannis2018mincoll} formalize the $k$-Dissimilar Paths with Minimum Collective Length ($k$DPML) problem where, given two road edges, they compute a set of $k$ paths containing sufficiently dissimilar routes and the lowest collective path length.
Hacker et al. \cite{hacker2021most} propose $k$-Most Diverse Near Shortest Paths (KMD) to recommend the set of $k$ near-shortest paths (based on a user-defined cost threshold) with the highest diversity (lowest pairwise dissimilarity). 
Dissimilarity approaches do not guarantee that a set of $k$ paths exists that satisfies the desired property.

\subsection*{Collective approaches}
In contrast with individual approaches, collective ones consider the impact of traffic in a collective environment where vehicles interact.
There are two main categories of collective approaches: one-shot and iterative methods.

\paragraph{One-shot methods.} 
They assign a route to each trip without further optimizing the routes. They are computationally efficient and provide a quick, yet not optimal, traffic allocation.
The simplest one-shot method is the All-Or-Nothing assignment (AON) \cite{campbell1950route}, in which each trip is assigned to the fastest path between the trip's origin and destination, considering the free-flow travel time.

Incremental Traffic Assignment (ITA) \cite{chen1991ita} extends AON incorporating the dynamic travel time changes within a road edge. 
ITA splits the mobility demand into $n$ splits of a specified percentage ($n = 4$ with 40\%, 30\%, 20\%, and 10\% are commonly used values \cite{wang2012understanding}).
The trips in the first split are assigned using AON, and then each edge's travel time is updated using the function proposed by the Bureau of
Public Roads (BPR) \cite{bpr1964traffic}. 
Next, the trips in the second split are assigned using AON, considering the updated travel time.
Iteratively, ITA assigns the trips in each split, updating the travel time at each iteration.

\paragraph{Iterative methods.} 
Iterative approaches employ multiple iterations to compute TA until a convergence criterion is satisfied. While these approaches can be computationally demanding, they offer the advantage of yielding the optimal solution once convergence is achieved.
Two main iterative approaches are the user equilibrium (UE) and the system optimum (SO).

UE is based on the Wardrop principle \cite{wardrop1952some}, which states that no individual driver can unilaterally improve their travel time by changing their route. 
In UE, each individual selfishly selects the most convenient path, and all the unused paths will have a travel time greater than the selected route. 
UE assumes that drivers are rational and have perfect network knowledge \cite{lujak2015route}.
However, a system in user equilibrium does not imply that the total travel time is minimized \cite{morandi2021bridging}. 
Dynamic User Equilibrium (DUE) \cite{friesz2010dynamic} approximates the user equilibrium by performing simulations to estimate travel times more accurately.

In contrast with UE, SO is based on Wardrop's second principle, which suggests drivers cooperate to minimize the total system travel time \cite{wardrop1952proceedings}. 
In SO, drivers are considered selfless and willingly adhere to assigned routes to reduce congestion and travel time.
Both UE and SO may be solved using an iterative algorithm for optimization. Beckmann et al. \cite{beckman1956studies} provide the mathematical models for the traffic assignment as a convex non-linear optimization problem with linear constraints that may be solved through an iterative algorithm to solve the quadratic optimization problems \cite{frank1956algorithm}.

One-shot methods are faster than iterative ones but offer only an approximation of the solution.  
Therefore, the choice between these approaches depends on the specific requirements of the problem, balancing accuracy with computational efficiency.

\subsubsection*{\bf Position of our Work}
METIS is a one-shot, cooperative approach that effectively and quickly solves TA by balancing environmental concerns and drivers' needs.

\section{METIS}
\label{sec:metis}
The idea behind METIS is to shift from an individualistic paradigm to a collective, cooperative one.\footnote{The name METIS has been inspired by the Greek goddess who personifies wisdom, cunning, strategy, and prudence.}
In contrast with existing AR algorithms, METIS acts as a central unit that provides drivers with suggested routes considering dynamic estimation of traffic conditions. 
METIS estimates vehicles' current positions to penalize edges expected to be traversed, thus avoiding congested edges.
Moreover, METIS incorporates a pattern-based choice criterion that discourages the selection of popular routes likely to be chosen by other drivers.  
By doing so, METIS optimizes the routing process and provides drivers with efficient paths that minimize travel time and alleviate traffic congestion.

Algorithm \ref{high_level_code} presents METIS' high-level pseudocode. 
It takes four inputs: \emph{(i)} a mobility demand $D$, i.e., a time-ordered collection of trips, each represented by its origin $o$, destination $d$, and departure time $t$; \emph{(ii)} a directed weighted graph $G=(V, E)$, representing the road network, where $V$ is the set of intersections and $E$ the set of road edges, each associated with the expected travel time estimated as its length divided by the maximum speed allowed; \emph{(iii)} a parameter $p > 0$, which controls to what extent to penalize crowded edges; \emph{(iv)} a slowdown parameter $s \geq 1$ accounting for reduced speeds on edges due to the presence of other vehicles and various events like traffic lights.

The algorithm starts with the initialization phase (lines 1-2), where it computes two $K_{\text{road}}$-based measures.
Subsequently, the algorithm performs the traffic assignment (lines 3-7): for each trip $j \in D$, METIS employs FLEP (Forward-Looking Edge Penalization) to penalize edges based on other vehicles' estimated current position, thus producing a penalized road network $H$ (line 4). 
Then, METIS employs KMD \cite{hacker2021most} to generate a set $P$ of $k$ alternative routes between each trip's origin $o$ and destination $d$, based on $H$ (line 5). 
Then, the algorithm assigns to the trip $j$ the route $r$ with the minimum value of a route scoring function (line 6), adding it to the routes collection $R$ (line 7). 
Once each trip in $D$ has been associated with a route, METIS returns $R$ (line 8).

The following sections provide details on METIS' components.
Section \ref{sec:kroads}  outlines the initialization phase and introduces the $K_{\text road}$-based measures,
Section \ref{sec:flep} introduces FLEP,
Section \ref{sec:kmdnsp} describes KMD, and Section \ref{sec:route_gen} describes route scoring.

\begin{algorithm}
    \SetKwInOut{Input}{Input}
    \SetKwInOut{Output}{Output}
    
    \Input{road network $G$, mobility demand $D$, penalization parameter $p$, slowdown parameter $s$}
    \Output{assigned routes $R$}
    
    \BlankLine
    \tcp{Initialization Phase}
    
    $K_{\text{road}}^{\text{\tiny (source)}}, K_{\text{road}}^{\text{\tiny (end)}} \leftarrow KRoadEstimation(G, D)$\;
    $R \leftarrow \emptyset$\;
    
    \BlankLine
    \tcp{Perform the Traffic Assignment (TA)}
    \ForEach{$j=(o, d, t)\in D$}{
    \BlankLine
        \tcp{Apply the Forward-Looking Edge Penalization (FLEP)}
        $H \leftarrow FLEP(G, R, p, s, t)$\;
        \BlankLine
        \tcp{Generate $k$ candidates on the penalized road network}
        $P \leftarrow KMD(H, o, d)$\;
        \BlankLine
        \tcp{Select the route that minimizes the route scoring function}
        $r\leftarrow RouteSelection(P, K_{\text{road}}^{\text{\tiny (source)}}, K_{\text{road}}^{\text{\tiny (end)}})$\;
        \BlankLine
        \tcp{Update the assigned routes collection}
        $R \leftarrow R \cup \{r\}$\;
    }
    \Return{$R$}\;
    \caption{METIS}
    \label{high_level_code}
\end{algorithm}


\subsection{Initialization Phase}
\label{sec:init}

During the initialization phase (line 1 of Algorithm \ref{high_level_code}), METIS calculates $K_{\text{road}}^{\text{\tiny(source)}}(e)$ and $K_{\text{road}}^{\text{\tiny(end)}}(e)$ for every edge $e$ in the road network. 
This computation requires a collection of routes to estimate sources and destinations of traffic on the road network. 
In contrast to the approach by Wang et al. \cite{wang2012understanding}, which utilizes real GPS data to compute $K_{\text{road}}$ for each edge, we adopt a more adaptable strategy. 
We establish connections between origin and destination points in $D$ with the fastest paths in the road network assuming free-flow travel time, enabling us to estimate the $K_{\text{road}}^{\text{\tiny(source)}}$ and $K_{\text{road}}^{\text{\tiny(end)}}$ values for each edge, even in situations where GPS data are unavailable.

\subsubsection*{\bf K$_{\text road}$ measures}
\label{sec:kroads}

The $K_{\text{road}}$ of an edge quantifies how many areas of the city (e.g., neighbourhoods) contribute to most of the traffic flow over that edge \cite{wang2012understanding}. 
The computation of $K_{\text{road}}$ involves constructing a road usage network, which is a bipartite network where each road edge is connected to its major driver areas, i.e., those responsible for 80\% of the traffic flow on that edge \cite{wang2012understanding}. 
The $K_{\text{road}}(e)$ of an edge $e$ is the degree of $e$ within the road usage network.
$K_{\text{road}}$ indicates an edge's popularity: an edge with a low $K_{\text{road}}$ is chosen by only a limited number of traffic sources, indicating relatively low popularity; an edge with a high $K_{\text{road}}$ attracts traffic from more diverse areas, indicating higher popularity among them.

We expand upon the $K_{\text{road}}$ concept by introducing $K_{\text{road}}^{\text{\tiny (source)}}$ and $K_{\text{road}}^{\text{\tiny (end)}}$ as follows. 
First, an area $A$ is a driver source for an edge $e$ if at least one vehicle originating from $A$ travels through $e$. 
Similarly, $A$ is a driver destination for $e$ if at least one vehicle traverses $e$ and completes its trip in $A$. 
An area can be both driver source and driver destination for a particular edge.
In this work, an area is a square tile of 1km within a square tessellation of the city.

We define the major driver sources (MDS) and the major driver destinations (MDD) as the areas to which 80\% of the traffic flowing through an edge starts or ends, respectively.
To calculate these two measures, we construct a bipartite network where a connection is established from an area $A$ to an edge $e$ if $A$ is an MDS for $e$. 
Similarly, a connection is formed from an edge $e$ to an area $A$ if $A$ is an MDD for $e$. 
Specifically, for an edge $e$, $K_{\text{road}}^{\text{\tiny (source)}}(e)$ is the in-degree of $e$ within the bipartite network, while $K_{\text{road}}^{\text{\tiny (end)}}(e)$ is $e$'s out-degree.

We also define $K_{\text{route}}^{\text{\tiny (source)}}(r)$ of a route $r = (e_1, \cdots, e_{n})$ as the average $K_{\text{road}}^{\text{\tiny (source)}}$
computed over its edges weighted with edge length:
\begin{equation}
K_{\text{route}}^{\text{\tiny (source)}}(r) = \frac{\sum_{i=1}^{n} K_{\text{road}}^{\text{\tiny (source)}}(e_i)\cdot l(e_i)}{\sum_{i=1}^{n} l(e_i)}
\end{equation}
\noindent where $l(e_i)$ is the length of edge $e_i$.
Similarly:
\begin{equation}
K_{\text{route}}^{\text{\tiny (end)}}(r) = \frac{\sum_{i=1}^{n} K_{\text{road}}^{\text{\tiny (end)}}(e_i)\cdot l(e_i)}{\sum_{i=1}^{n} l(e_i)}
\end{equation}

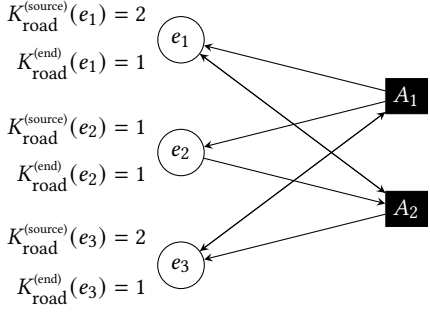
\begin{figure}
    \centering
\begin{tikzpicture}
  \node[draw, circle] (e1) at (0,0) {$e_1$};
  \node[draw, circle] (e2) at (0,-1.5) {$e_2$};
  \node[draw, circle] (e3) at (0,-3) {$e_3$};
  
  \node[draw, rectangle, fill=black, text=white] (B1) at (3,-0.75) {$A_1$};
  \node[draw, rectangle, fill=black, text=white] (B2) at (3,-2.25) {$A_2$};

  \node[left, xshift=-10pt, yshift=+9pt] at (e1) {$K_{\text{road}}^{\text{\tiny (source)}}(e_1) = 2$};
  \node[left, xshift=-10pt, yshift=-9pt] at (e1) {$K_{\text{road}}^{\text{\tiny (end)}}(e_1) = 1$};
  
  \node[left, xshift=-10pt, yshift=+9pt] at (e2) {$K_{\text{road}}^{\text{\tiny (source)}}(e_2) = 1$};
  \node[left, xshift=-10pt, yshift=-9pt] at (e2) {$K_{\text{road}}^{\text{\tiny (end)}}(e_2) = 1$};
  
  \node[left, xshift=-10pt, yshift=+10pt] at (e3) {$K_{\text{road}}^{\text{\tiny (source)}}(e_3) = 2$};
  \node[left, xshift=-10pt, yshift=-9pt] at (e3) {$K_{\text{road}}^{\text{\tiny (end)}}(e_3) = 1$};
  
  
  \draw[->, >=stealth] (e1) -- (B2);
  \draw[->, >=stealth] (e2) -- (B2);
  \draw[->, >=stealth] (e3) -- (B1);

  \draw[->, >=stealth] (B1) -- (e2);
  \draw[->, >=stealth] (B1) -- (e3);
  \draw[->, >=stealth] (B2) -- (e3);
  \draw[->, >=stealth] (B1) -- (e1);
  \draw[->, >=stealth] (B2) -- (e1);
  
\end{tikzpicture}
\caption{Graphical representation of the bipartite network of road edges and areas. 
$K_{\text{road}}^{\text{\tiny (source)}}$ is the in-degree of edge nodes, $K_{\text{road}}^{\text{\tiny (end)}}$ is the out-degree of edge nodes.}
\label{fig:kroad_schema}
\end{figure}

\noindent \textbf{Example.} Figure \ref{fig:kroad_schema} illustrates the concepts of $K_{\text{road}}^{\text{\tiny (source)}}$ and $K_{\text{road}}^{\text{\tiny (end)}}$ with three edges ($e_1, e_2, e_3$, circles) connected with two areas ($A_1, A_2$, squares).
Let us consider edge $e_1$: it has one outgoing connection towards $A_2$, leading to an out-degree of 1, and thus $K_{\text{road}}^{\text{\tiny (end)}}(e_1) = 1$.
Moreover, edge $e_1$ has
incoming connections from areas $A_1$ and $A_2$, resulting in an in-degree of 2 and, consequently, $K_{\text{road}}^{\text{\tiny (source)}}(e_1) = 2$.

\subsection{Forward-Looking Edge Penalization}
\label{sec:flep}

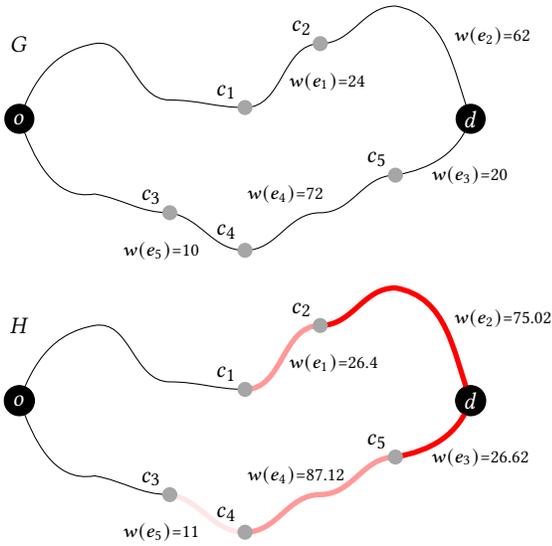
\begin{figure}
\centering

\begin{tikzpicture}

\draw[] (0,0) to[out=70,in=190] (1,1) to[out=10,in=180] (2,0.25) to[out=0,in=180] (3,0.15) to[out=0,in=180] (4,1) to[out=0,in=180] (5,1.5) to[out=-10,in=110] (6,0);

\draw[] (0,0) to[out=-70,in=190] (1,-1) to[out=-10,in=180] (2,-1.25) to[out=-10,in=180] (3,-1.75) to[out=0,in=180] (4,-1.25) to[out=0,in=180] (5,-0.75) to[out=10,in=-110] (6,0);

\draw[fill=black] (0,-3.75) circle [radius=0.2];
\draw[fill=black] (6,-3.75) circle [radius=0.2];

\draw[] (0,-3.75) to[out=70,in=190] (1,-2.75) to[out=10,in=180] (2,-3.5) to[out=0,in=180] (3,-3.6) to[out=0,in=180] (4,-2.75) to[out=0,in=180] (5,-2.25) to[out=-10,in=110] (6,-3.75);

\draw[color=red!40, line width=2pt] (3,-3.6) to[out=0,in=180] (4,-2.75);

\draw[color=red, line width=2pt] (4,-2.75) to[out=0,in=180] (5,-2.25) to[out=-10,in=110] (6,-3.75);

\draw[] (0,-3.75) to[out=-70,in=190] (1,-4.75) to[out=-10,in=180] (2,-5) to[out=-10,in=180] (3,-5.5) to[out=0,in=180] (4,-5) to[out=0,in=180] (5,-4.5) to[out=10,in=-110] (6,-3.75);

\draw[color=red!10, line width=2pt] (2,-5) to[out=-10,in=180] (3,-5.5);

\draw[color=red!40, line width=2pt] (3,-5.5) to[out=0,in=180] (4,-5) to[out=0,in=180] (5,-4.5);

\draw[color=red, line width=2pt] (5,-4.5) to[out=10,in=-110] (6,-3.75);


\node[right] at (3.457, 0.5) {\footnotesize{$w(e_1)$=24}};
\node[right] at (5.65, 1.1) {\footnotesize{$w(e_2)$=62}};
\node[right] at (5.35, -0.75) {\footnotesize{$w(e_3)$=20}};
\node[right] at (2.9,-1) {\footnotesize{$w(e_4)$=72}}; 
\node[right] at (1.25,-1.75) {\footnotesize{$w(e_5)$=10}};

\node[right] at (3.457, -3.25) {\footnotesize{$w(e_1)$=26.4}};  
\node[right] at (5.65, -2.65) {\footnotesize{$w(e_2)$=75.02}};
\node[right] at (5.35, -4.5) {\footnotesize{$w(e_3)$=26.62}};    
\node[right] at (2.9,-4.75) {\footnotesize{$w(e_4)$=87.12}};     
\node[right] at (1.25,-5.5) {\footnotesize{$w(e_5)$=11}};


\fill[gray!70] (3,0.15) circle [radius=0.1];
\fill[gray!70] (4,1) circle [radius=0.1];
\node[left, xshift=0pt, yshift=+6pt] at (3,0.15) {$c_1$};
\node[left, xshift=0pt, yshift=+6pt] at (4,1) {$c_2$};
\node[left, xshift=0pt, yshift=+6pt] at (2,-1.25) {$c_3$};
\node[left, xshift=0pt, yshift=+7pt] at (3,-1.75) {$c_4$};
\node[left, xshift=0pt, yshift=+6pt] at (5,-0.75) {$c_5$};

\fill[gray!70] (3,-3.6) circle [radius=0.1];
\fill[gray!70] (4,-2.75) circle [radius=0.1];

\node[left, xshift=0pt, yshift=+6pt] at (3,-3.6) {$c_1$};
\node[left, xshift=0pt, yshift=+6pt] at (4,-2.75) {$c_2$};

\fill[gray!70] (2,-5) circle [radius=0.1];
\fill[gray!70] (3,-5.5) circle [radius=0.1];
\fill[gray!70] (5,-4.5) circle [radius=0.1];
\node[left, xshift=0pt, yshift=+6pt] at (2,-5) {$c_3$};
\node[left, xshift=0pt, yshift=+7pt] at (3,-5.5) {$c_4$};
\node[left, xshift=0pt, yshift=+6pt] at (5,-4.5) {$c_5$};

\node[] at (0, 1) {$G$};
\node[] at (0, -2.75) {$H$};

\fill[gray!70] (2,-1.25) circle [radius=0.1];
\fill[gray!70] (3,-1.75) circle [radius=0.1];
\fill[gray!70] (5,-0.75) circle [radius=0.1];

\node[circle, color=white, fill=black, inner sep=0pt, minimum size=0.4cm] at (0,0) {$o$};
\node[circle, color=white, fill=black, inner sep=0pt, minimum size=0.4cm] at (6,0) {$d$};

\node[circle, color=white, fill=black, inner sep=0pt, minimum size=0.4cm] at (0,-3.75) {$o$};
\node[circle, color=white, fill=black, inner sep=0pt, minimum size=0.4cm] at (6,-3.75) {$d$};

\end{tikzpicture}
\caption{FLEP with $p=0.1$ applied to road network $G$, resulting in penalized network $H$. Grey circles represent estimated vehicle positions. FLEP applies cumulative penalization to edges based on the vehicles' expected traversal, with a multiplicative factor of $(1+p)$. Darker red color indicates higher penalties imposed on road edges. For example, edge $e_4$ is traversed by vehicles $c_3$ and $c_4$, leading to a penalty of $(1+p)^2$, resulting in $w(e_4) = 72 \cdot (1.1)^2 = 87.12$.}
\label{fig:flep}
\end{figure}

 Forward-Looking Edge Penalization (FLEP) is based on penalizing road edges to reflect the dynamic changes in travel time caused by increasing traffic volume.

Generally, existing methods penalize the entire routes assigned to currently travelling vehicles \cite{chen1991ita, hacker2021most, liu2018finding}. 
However, this indiscriminate penalization of all edges, including those currently unoccupied, may discourage the utilization of potentially efficient routes, leading to congestion in alternative paths that are not penalized. 

FLEP overcomes this problem by estimating the current positions of vehicles in transit and penalizing the edges that these vehicles are projected to visit.   
Assuming that a vehicle departed $t$ seconds ago, FLEP computes the distance it has travelled during $t$ seconds, assuming that the vehicle travelled at a speed of $max\_speed/s$ on each edge, where $s$ is a slowdown parameter accounting for reduced speeds on edges due to the presence of other vehicles and various events like traffic lights. 
Then, FLEP modifies the weights $w(e)$ assigned to the edges that the vehicle is expected to traverse by applying a penalty factor $p$: $w(e) \leftarrow w(e)\cdot (1+p)$. The penalization is cumulative, i.e., the edge is penalized for each vehicle that will traverse that edge.
This penalization discourages the selection of edges that vehicles are likely to traverse, promoting alternative routes and a balanced distribution of traffic.

Algorithm \ref{high_level_code_flep} provides the pseudocode of FLEP.
First, FLEP considers each previously assigned route $r$ and calculates the time $\Delta t$ the vehicle spent travelling based on its departure time $t(r)$ and the current time $t$ (line 2).
Then, it computes the required travel time to reach each edge $e \in r$ using $s$ (line 3). 
If the vehicle has yet to reach its destination (line 4), FLEP determines the index of the first unvisited edge in the route (line 5). 
Subsequently, it penalizes every unvisited edge in route $r$ (lines 7-8). Finally, FLEP outputs the penalized network (line 9).

\textbf{Example.} Figure \ref{fig:flep} illustrates how FLEP works, assuming a penalization $p=0.1$. 
FLEP estimates the position of each vehicle in transit (grey circles) within the road network considering $s$. 
Subsequently, FLEP applies cumulative penalization to the edges that each vehicle will traverse to reach the destination $d$. 
This penalization is accomplished by multiplying the weights of these edges by $(1+p)$ for each vehicle that will traverse it.
For example, vehicles $c_3$ and $c_4$ are projected to pass through edge $e_4$. 
Consequently, the initial weight $w(e_4) = 72$ is penalized by $(1+p)^2$, resulting in a new weight of $w(e_4) = 72 \cdot (1.1)^2 = 87.12$.
FLEP generates a modified road network $H$ through this iterative process, penalizing edges according to the anticipated vehicle movements.

\begin{algorithm}
    \SetKwInOut{Input}{Input}
    \SetKwInOut{Output}{Output}
    
    \Input{road network $G$, collection of assigned paths $R$, penalty parameter $p$, slowdown parameter $s$, current time $t$}
    \Output{updated road network $G$}

    \BlankLine
    \ForEach{$r\in R$}{
    \BlankLine
        \tcp{Compute the vehicle's time spent travelling. $t(r)$ is the departure time of $r$}
        $\Delta t = t - t(r)$\;
        \BlankLine
        \tcp{list of travel times}
        $tt\_list \leftarrow TravelTimesComputation(r, s)$\;
    \BlankLine
    \tcp{Penalize unvisited edges}
    \If{$\Delta t < tt\_list[len(r)]$}{
        $i \leftarrow \min \{x \in [1, len(r)] | tt\_list[x]>\Delta t\}$\;
        \For{$j \in [i, \dots, len(r)]$}{
        $ e \leftarrow r [j]$\;
            $w(e) \leftarrow w(e)\cdot (1+p)$\;
        }    
    }   
    }
    \tcp{Return the road network with the penalized weights}
    \Return{$G$}\;

     \BlankLine
     \SetKwFunction{TravelTimesComputation}{$TravelTimesComputation$}
    \SetKwProg{Fn}{Function}{:}{}
    \Fn{\TravelTimesComputation{$r, s$}}{
        $tt_{agg} \leftarrow 0$\;
        $l \leftarrow []$\;
        \ForEach{$e\in r$}{
            $tt_{agg} \leftarrow tt_{agg} + w(e)\cdot s$ \;
            $l.append(tt_{agg})$
    }
    \Return{$l$}\;
}

    \caption{Forward-Looking Edge Penalization (FLEP)}
    \label{high_level_code_flep}
\end{algorithm}

\subsection{KMD}
\label{sec:kmdnsp}

$k$-Most Diverse Near Shortest Paths (KMD) is an AR algorithm that generates a collection of $k$ routes with the highest dissimilarity among each other while still adhering to a user-defined cost threshold $\epsilon$ \cite{hacker2021most}. 
As KMD becomes computationally challenging for $k > 2$ due to its NP-hard nature, a penalization-based heuristic is commonly employed to accelerate the computation process \cite{hacker2021most}.

Given an origin $o$ and a destination $d$, KMD first calculates the fastest path between $o$ and $d$. 
The cost $c$ of this path, along with the parameter $\epsilon$, determines the maximum allowed cost threshold $c \cdot (1+\epsilon)$ for a path to be considered near-shortest.
Next, KMD iteratively applies the penalization-based heuristic to compute a new near-shortest path $p$, which is then added to the set of near-shortest paths $S$. Subsequently, it generates all subsets of $S$ composed of $k$ elements. 
Among these subsets, KMD identifies the most diverse using the Jaccard coefficient, which compares the dissimilarity between pairs of paths.
When no more near-shortest paths can be found using the penalization approach, KMD returns the subset of $k$ paths with the highest diversity.
The detailed pseudo-code of KMD is in \cite{hacker2021most}.

In this study, we use parameter values $k = 3$ (three alternative routes) and $\epsilon = 0.3$ (maximum cost increase of 30\% compared to the fastest path) for KMD. 
These parameter values are commonly used in alternative routing algorithms \cite{li2022comparing, hacker2021most}.

\subsection{Route Selection}
\label{sec:route_gen}

In the final step, METIS scores and ranks the set of alternative routes generated by KMD. 
To determine the best route among the alternatives, METIS assigns a score (the lower, the better) to each route $r$ based on the following formula:
\begin{equation}
score(r) = \frac{K_{\text{road}}^{\text{\tiny(source)}}(r) \cdot K_{\text{road}}^{\text{\tiny(end)}}(r) }{C_r}
\label{eq:score}
\end{equation}
\noindent where $C_r$ is the average of the capacities $C(e)$ of the edges in route $r$, taking into account the edge length. 
The capacity $C(e)$ of an edge $e$ is computed as follows: 
\begin{equation}
C(e)=\begin{cases} 1900 \cdot l \cdot q & \mbox{if }s_{\text max}\leq45 \\
(1200+20\cdot s_{\text max}) \cdot l & \mbox{if } 45 < s_{\text max} <60  \\
(1700+10\cdot s_{\text max}) \cdot l & \mbox{if } s_{\text max} \geq 60
\end{cases}
\end{equation}
\noindent where $s_{\text max}$ is the speed limit associated with edge $e$ (in miles/hour), $l$ is the number of lanes in edge $e$, and $q=0.5$ is the green time-to-cycle length ratio. 
The equation and the values above are taken from the 2000 Highway Capacity Manual \cite{highwaycapacitymanual2000, wang2012understanding}. 

Route scoring combines two essential elements. 
In the denominator, the average capacity $C_r$ favours routes composed mainly of high-capacity edges, which are expected to handle larger traffic volumes. 
In the numerator, the product $K_{\text{road}}^{\text{\tiny(source)}}(r) \cdot K_{\text{road}}^{\text{\tiny(end)}}(r)$ penalizes routes that predominantly consist of popular edges, promoting a balanced traffic distribution.

\section{Experimental Setup}
\label{sec:experimental_setup}

This section describes the experimental settings employed in our study (Section \ref{sec:exp_settings}), an overview of the baselines we compare with METIS (Section \ref{sec:baselines}), and the measures used for the comparison (Section \ref{sec:measures}).

\subsection{Experimental Settings}
\label{sec:exp_settings}

We conduct experiments in three Italian cities: Milan, Rome, and Florence. 
These cities represented diverse urban environments with varying traffic dynamics, sizes, and road networks (Table \ref{tab:road_networks}).

\subsubsection*{\bf Road Networks}
We obtain a road network for each city using OSMWebWizard.
The three cities' road network characteristics are heterogeneous (see Table \ref{tab:road_networks}). 
While the smallest city, Florence's road network exhibits the highest density (9.11).
Milan and Rome are sparse compared to Florence, although they have extensive road networks. 
This difference in road network characteristics provides a valuable basis for evaluating the performance of TA algorithms in different urban contexts.

\subsubsection*{\bf Mobility Demand}
\label{sec:mobility_demand}
We split each city into 1km squared tiles using a GPS dataset provided by Octo \cite{scikitmob, bohm2022gross, giannotti2011unveiling} to determine each vehicle's trip's starting and ending tiles. 
We use this information to create an origin-destination matrix $M$, where $m_{o, d}$ represents the number of trips starting in tile $o$ and ending in tile $d$.
To generate a mobility demand $D$ of $N$ trips, we randomly select a trip $T_v=(e_o, e_d)$ for a vehicle $v$, choosing matrix elements $m_{o, d}$ with probabilities $p_{o, d} \propto m_{o, d}$. 
We then uniformly select two edges $e_o$ and $e_d$ within tiles $o$ and $d$ from the road network $G$.
For our experiments, we set $N=10$k trips in Florence, $N=20$k trips in Rome, and $N=30$k in Milan. 
These values are chosen to minimize the difference between the travel time distribution of GPS trajectories and those obtained from a simulation of a rush hour in SUMO, a standard method for assessing the realism of simulated traffic \cite{cornacchia2022how, argota2022getting}.

\begin{table}[htb]
\centering
\begin{tabular}{@{}lrrr|rrr|r@{}}
\toprule
\bf city & \bf $|V|$ & \bf $|E|$ & \bf $L(E)$ & \bf area & \bf den & \bf \# trips & $N$\\ 
\midrule
Florence & 6,140 & 11,804 & 1,050 & 115.28 & 9.11  & 4,076 & 10k\\
Milan & 24,063 & 46,488 & 4,340 & 495.55 & 8.76 & 5,617 & 30k \\
Rome & 31,798 & 63,384 & 6,569 & 788.11 & 8.34  & 7,622 & 20k\\ \bottomrule
\end{tabular}
\caption{Overview of the road network characteristics in the three cities. 
The columns show the number of vertices $|V|$ and edges $|E|$, the total road length $L(E)$, the area of the city, the ratio of road length to surface area (den, in km road/km$^2$), the total number of trips described by GPS data, and the number $N$ of routes generated in each city.}
\label{tab:road_networks}
\end{table}

\subsection{Baselines}
\label{sec:baselines}
We evaluate METIS against several one-shot TA solutions, both individual and collective. 
We exclude iterative solutions like User Equilibrium (UE) \cite{wardrop1952some, friesz2010dynamic} and System Optimum (SO) \cite{wardrop1952proceedings} from our analysis. 
While these iterative approaches may offer optimal results after convergence, their computationally intensive nature and multiple iterations make them unsuitable for real-time applications.

\subsubsection*{\bf AR baselines}
AR algorithms are designed to generate $k$ alternative routes for an individual trip. 
We extend these algorithms to TA by aggregating the recommended routes for each trip within a mobility demand.
In particular, we use an AR algorithm to compute $k = 3$ alternative routes for each trip in mobility demand $D$, and we randomly select one of them uniformly.
In this study, we consider the following state-of-the-art methods:
\begin{itemize}
\item \textbf{PP (Path Penalization)} generates $k$ alternative routes by penalizing the weights of edges contributing to the fastest path \cite{cheng2019shortest}. In each iteration, PP computes the fastest path and increases the weights of the edges that contributed to it by a factor $p$ as $w(e) = w(e)\cdot(1+p)$. 
The
penalization is cumulative: if an edge has already been penalized in a previous iteration, its weight will be further increased \cite{cheng2019shortest}. 

\item \textbf{GR (Graph Randomization)} generates $k$ alternative paths by randomizing the weights of all edges in the road network before each fastest path computation. The randomization is done by adding a value from a normal distribution, given by the equation $N(0, w(e)^2\cdot \delta^2)$ \cite{cheng2019shortest}. 

\item \textbf{PR (Path Randomization)} generates $k$ alternative paths randomizing only the weights of the edges that were part of the previously computed path.
Similar to GR, it adds a value from a normal distribution to the edge weights, following the equation $N(0, w(e)^2\cdot \delta^2)$ \cite{cheng2019shortest}. 

\item \textbf{KD ($k$-shortest disjointed paths)} returns $k$ alternative non-overlapping paths (i.e., with no common edges) \cite{suurballe1974disjoint}.

\item \textbf{PLA (Plateau)} builds two shortest-path trees, one from the origin and one from the destination, and identifies their common branches (plateaus) \cite{camvit2005choice}.
The top-$k$ plateaus are selected based on their lengths, and alternative paths are generated by appending the fastest paths from the source to the plateau's first edge and from the last edge to the target.

\item \textbf{KMD ($k$-Most Diverse Near Shortest Paths)} generates $k$ alternative paths with the highest dissimilarity among each other while adhering to a user-defined cost threshold $\epsilon$ \cite{hacker2021most}. 

\end{itemize}

\subsubsection*{\bf One-shot baselines}
In contrast with AR approaches, one-shot (OS) ones assign a route to each trip of a mobility demand without further optimization on the assigned routes. 
In this study, we consider the two most common OS approaches:
\begin{itemize}
    \item \textbf{AON (All-Or-Nothing)} assigns each trip to the fastest path between the trip's origin and destination, assuming free-flow travel times \cite{campbell1950route}.
    \item \textbf{ITA (Incremental Traffic Assignment)} \cite{chen1991ita} uses four splits (40\%, 30\%, 20\%, 10\%, as recommended in the literature \cite{wang2012understanding}) to assign routes to trips. 
    In the first split, ITA uses AON considering free-flow travel time $t_{\text{\tiny free}}$. 
    It then updates the travel times using the Bureau of Public Roads (BPR) function $t_a = t_{\text{\tiny free}} \cdot(1+\alpha \cdot \mbox{VOC}^\beta)$ \cite{bpr1964traffic}, where $\mbox{VOC}$ indicates an edge's traffic volume over its capacity and $\alpha=0.15$ and $\beta=4$ are values recommended in the literature \cite{wang2012understanding, morandi2021bridging}. 
This process is repeated for each split, progressively updating the travel times and assigning trips accordingly.
\end{itemize}

Table \ref{tab:best_params} shows the parameter ranges tested for each baseline and the best parameter combinations obtained in our experiments.

\begin{table}[htb]
\centering
\begin{tabular}{@{}lllll@{}}
\toprule
 & & \multicolumn{3}{c}{\bf best params} \\
 \cline{3-5}
\bf algo & \bf params range & \bf Florence & \bf Milan & \bf Rome \\ \midrule
PP & \footnotesize{$p \in \{.1, .2, \dots, .5\}$} & $p=.2$ & $p=.1$ & $p=.2$ \\
GR & \footnotesize{$\Delta \in \{.2, .3, .4, .5\}$} & $\Delta=.2$ & $\Delta=.2$ & $\Delta=.2$ \\
PR & \footnotesize{$\delta \in \{.2, .3, .4, .5\}$} & $\Delta=.2$ & $\Delta=.2$ & $\Delta=.3$ \\
KMD & \footnotesize{$\epsilon \in \{.01, .05, .1, .2, .3\}$} & $\epsilon=.2$ & $\epsilon=.1$ & $\epsilon=.01$ \\
\hline
\multirow{2}{*}{METIS} & \footnotesize{$p \in \{.01, .015, .02, \dots, .1\}$} & $p=.025$ & $p=.01$ & $p=0.01$ \\ 
& \footnotesize{$s \in \{1.5, 1.75, 2, 2.25\}$} & $s=2.25$ & $s=2.25$ & $s=1.75$ \\
\bottomrule
\end{tabular}
\caption{Parameter values explored for each algorithm and the best values obtained for each approach.}
\label{tab:best_params}
\end{table}

\subsection{Measures}
\label{sec:measures}
To assess the effectiveness of METIS and the baselines, we use three measures: total CO2 emissions, road coverage, and redundancy. 

\subsubsection*{\bf Total CO2}
To accurately account for vehicle interactions and calculate CO2 emissions, we utilize the traffic simulator SUMO (Simulation of Urban MObility) \cite{Microscopic2018Lopez, cornacchia2022how}, which simulates each vehicle's dynamics, considering interactions with other vehicles, traffic jams, queues at traffic lights, and slowdowns caused by heavy traffic. 

For each city and algorithm, we generate $N$ routes (with $N$ depending on the city, see Table \ref{tab:road_networks}) and simulate their interaction within SUMO during one peak hour, uniformly selecting a route's starting time during the hour.

To estimate CO2 emissions related to the trajectories produced by the simulation, we use the HBEFA3 emission model ~\cite{infras2013handbook, bohm2022gross}, which estimates the vehicle's instantaneous CO2 emissions at a trajectory point $j$ as:
\begin{equation}
\mathcal{E}(j) = c_0 + c_1sa + c_2sa^2 + c_3s + c_4s^2 + c_5s^3 
\end{equation}
where $s$ and $a$ are the vehicle's speed and acceleration in point $j$, respectively, and $c_0,\dots,c_5$ are parameters changing per emission type and vehicle taken from the HBEFA database ~\cite{krajzewicz2015second}.
To obtain the total CO2 emissions, we sum the emissions corresponding to each trajectory point of all vehicles in the simulation. 

\subsubsection*{\bf Road Coverage (RC)}
It quantifies the extent to which the road network is utilized by vehicles. 
It is calculated by dividing the total distance vehicles cover on visited edges by the road network's overall length.
Mathematically, given a set of routes $R$ and the set of edges in these routes $S_R = \bigcup_{r \in R} \{ e \in r\}$, we define RC as:
\begin{equation}
RC(R) = \frac{\sum_{e \in S_R} l(e)}{L(E)} \cdot 100
\end{equation}
\noindent where $l(e)$ is the length of edge $e$ and $L(E)=\sum_{e \in E} l(e)$ is the total road length of the road network.

Road coverage characterizes a TA algorithm's road infrastructure usage. 
A higher road coverage indicates a larger proportion of the road network being utilized, which typically results in improved traffic distribution and reduced congestion. 
However, excessively high road coverage may increase vehicle travel distances, potentially producing higher emissions. 
Therefore, road coverage is a critical metric for evaluating the effectiveness of TA algorithms in effectively utilizing road infrastructure.

\subsubsection*{\bf Time redundancy (RED)} 
In the literature, redundancy is defined as the popularity of edges in a set of routes, also interpreted as the average utilization of edges that appear in at least one route \cite{cheng2019shortest}. 
Specifically, it is the fraction of the total number of edges of all routes divided by the total number of unique edges of all routes. 
Formally, given a set of routes $R$ and a set of edges in these routes $S_R = \bigcup_{r \in R} \{ e \in r\}$, we define it as:
\begin{equation}
RED(R) = \frac{\sum_{r \in R} |r|}{|S_R|}
\end{equation}
If $RED(R)=1$, there is no overlap among the routes in $R$, while $RED(R)=|R|$ when all routes are identical.

Note that RED does not consider traffic's dynamic evolution. 
To account for it, we define time redundancy as:
\begin{equation}
RED(R, t) = \frac{1}{|I|}\sum_{i\in I} \text{RED}(R_{i, t})
\end{equation}
where $t$ is the length of the time window, $I=\{t_0, t_0 + \sigma, t_0 + 2\sigma, \dots, t_{max}\}$ is the set of the starting times of each time window in the observation period $[t_0, t_{max}$) shifted by $\sigma$, and $\text{RED}(R_{i, t})$ is the RED of trips in $R$ departed within time interval $[i, i+t)$.
Low $RED(R, t)$ indicates that routes close in time are better distributed across edges.

\section{Results}
\label{sec:results}

\begin{figure*}
    \centering    \includegraphics[width=0.9\textwidth]{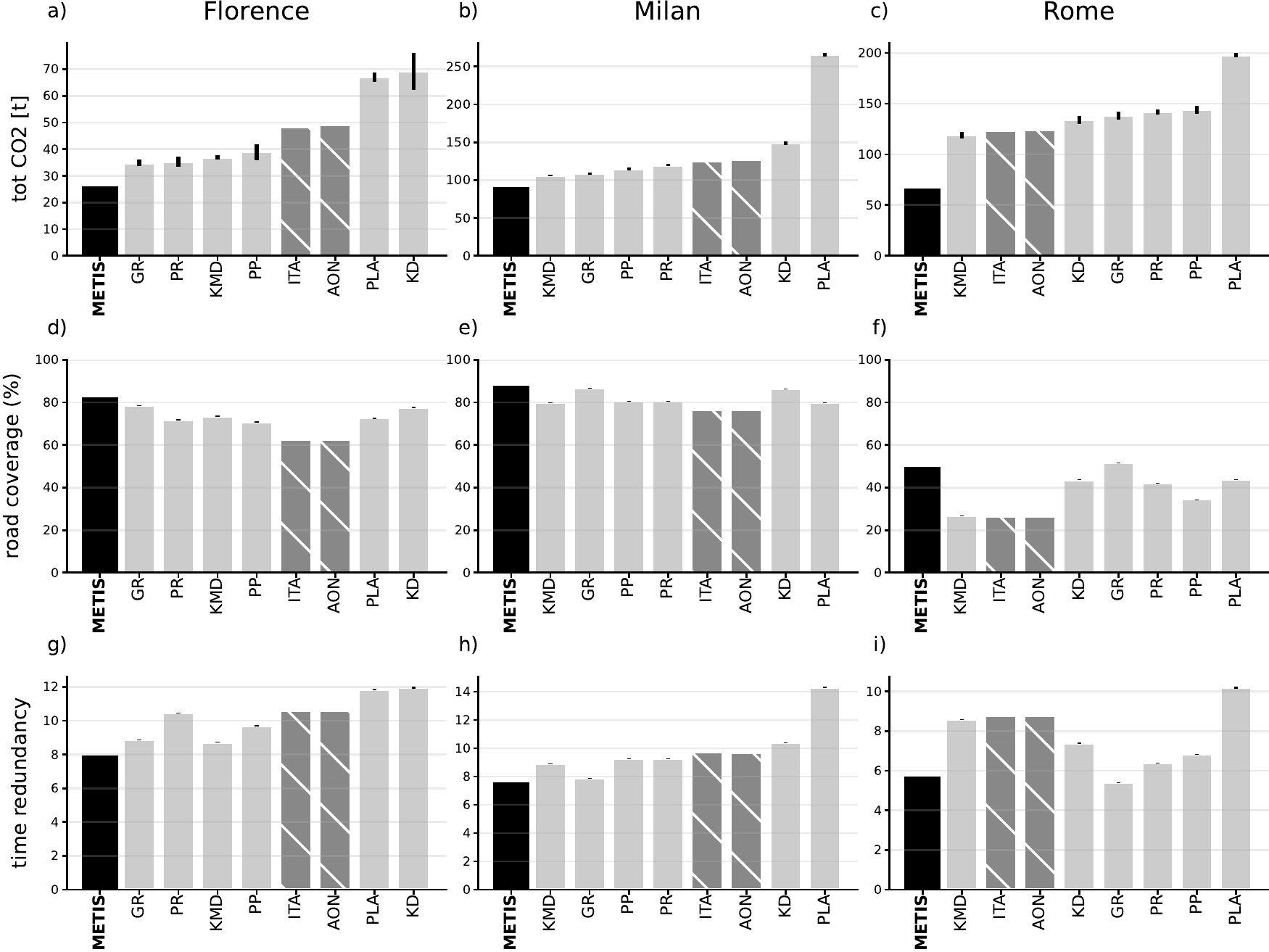}
    \caption{Comparison of METIS (black bar) with the baselines in Florence, Milan, and Rome on CO2 emissions (in tons), road coverage (in \%), and time redundancy. 
    To ensure statistical reliability, we run non-deterministic algorithms (GR, PR, KMD, PP, PLA, KD) ten times and present the average values and the corresponding standard deviation.
}
    \label{fig:results}
\end{figure*}

Table \ref{tab:results} and Figure \ref{fig:results} compare METIS with all the baselines for all cities and measures. 
For each model, we show the results regarding the combination of parameter values leading to the lowest CO2 emissions (see Table \ref{tab:best_params}).

METIS emerges as a significant breakthrough, with impressive reductions of CO2 emissions of 28\% in Florence, 18\% in Milan, and 46\% in Rome compared to the best baseline (see Figure \ref{fig:results}a-c and Table \ref{tab:results}). 
This remarkable result is due to the synergistic combination of its unique core components: FLEP, KMD, and route scoring.
FLEP is crucial in identifying less congested routes by estimating vehicles' current positions and dynamically adjusting edge weights.
Complementing FLEP, KMD offers alternative routes that substantially cover the road network. 
Lastly, route scoring prioritizes less popular routes with higher capacity, helping accommodate traffic volume over uncongested routes.

Indeed, METIS achieves the highest road coverage in Florence (79.66\%) and Milan (86.68\%) and the second-highest in Rome (48.51\%) (see Figure \ref{fig:results}d-f and Table \ref{tab:results}). 
Moreover, METIS achieves the lowest time redundancy in Florence (7.81) and Milan (7.41) and the second lowest in Rome (5.57): on average, the number of routes on each edge within a 5-minute temporal window is relatively low. 

\begin{figure}[htb]
    \centering
\subfigure[\large KMD]{
\includegraphics[width=0.75\columnwidth]{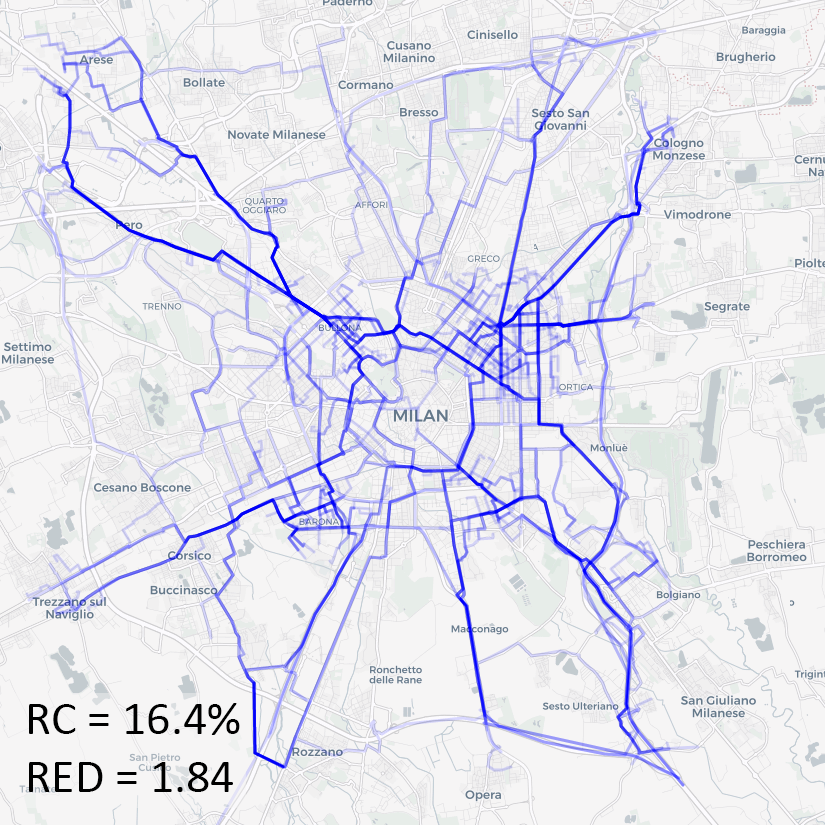}{}}
\subfigure[\large METIS]{
\includegraphics[width=0.75\columnwidth]{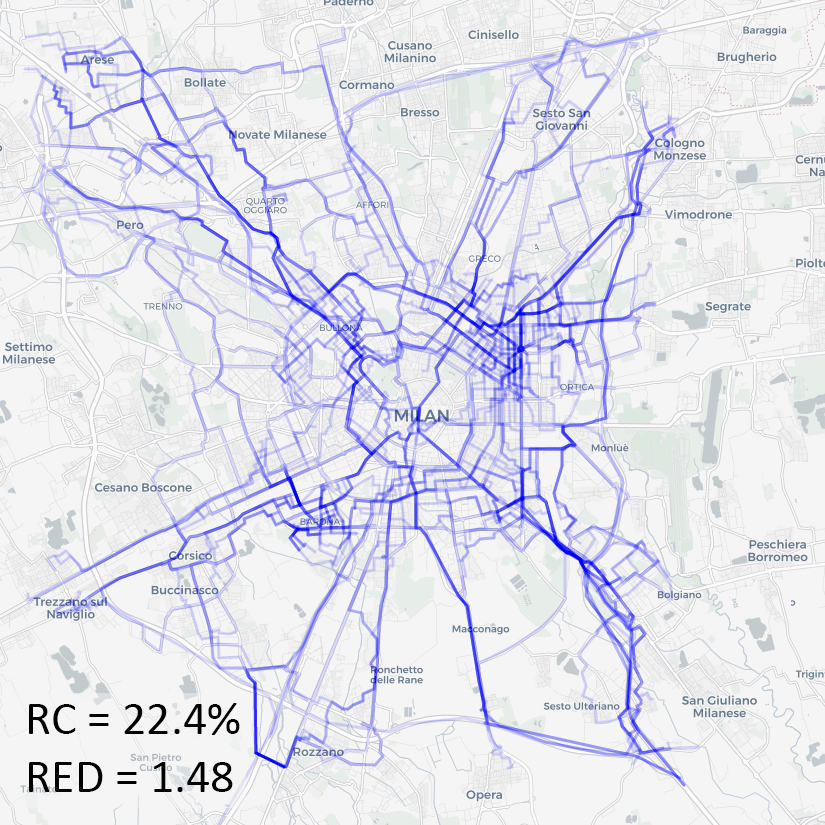}}
    \vspace{-0.4cm}
    \caption{Routes generated by KMD (a) and METIS (b) in Milan for 150 trips. 
    METIS exhibits a more spatially uniform distribution of traffic than KMD, which tends to concentrate routes on highly popular routes.
    RC indicates the road coverage and RED the time redundancy (5-minute window).}
    \label{fig:map}
\end{figure}

Figure \ref{fig:map} visually illustrates the spatial distribution of sample routes generated by METIS and KMD (the second-best model) in Milan. It is evident from the figure that METIS produces routes that are more evenly distributed across the city, leading to higher road coverage and lower time redundancy compared to KMD.

Among the baselines, GR shows the lowest CO2 emissions in Florence, while KMD is the best in Milan and Rome. 
GR has a high road coverage of 78.35\% in Florence, 86.57\% in Milan, and 51.57\% in Rome (see Figure \ref{fig:results}d-f and Table \ref{tab:results}). In Rome, GR achieves a higher road coverage than METIS.

\begin{table}[htb]
  \centering
  \small
  \begin{tabular}{c | c | lccc}
    \toprule
    \multicolumn{1}{c}{} & \multicolumn{1}{c}{} & \textbf{algo} & \textbf{CO2 [t]} & \textbf{RC (\%)} &  \textbf{RED$(R, 5m)$} \\
    \midrule
    \multirow{9}{*}{\rotatebox[origin=c]{90}{Florence}} & \multirow{6}{*}{ \rotatebox[origin=c]{90}{\small AR}} & PP & 38.94 {\small (2.99)} & 70.83 {\small(.30)} &  9.70 {\small(.03)} \\
    & & GR & 34.78 {\small(1.21)} & 78.35 {\small(.26)} & 8.87 {\small(.03)} \\
    & & PR & 35.20 {\small(1.88)} & 71.65 {\small(.37)} & 10.45 {\small(.03)} \\
    & & KD & 69.13 {\small(6.96)} & 77.48 {\small(.30)} & 11.94 {\small(.06)} \\
    & & PLA & 67.00 {\small(1.67)} & 72.56 {\small(.37)} & 11.83 	{\small(.04)} \\
    & & KMD & 36.93 {\small(.93)} & 73.40 {\small(.35)} & 8.72 	{\small(.03)} \\
    \cline{2-6}
    & \multirow{2}{*}{ \rotatebox[origin=c]{90}{\small OS}} & AON & 49.02 & 62.41 & 10.58\\
    & & ITA & 48.16 & 62.42 & 10.58 \\
    \cline{2-6}
    \rowcolor{lavender}
    \multicolumn{1}{c}{} & \multicolumn{1}{c}{} & \bf METIS &  \textbf{25.19} & \textbf{81.06} & \textbf{7.81} \\
    \midrule
    
    \multirow{9}{*}{\rotatebox[origin=c]{90}{Milan}} & \multirow{6}{*}{ \rotatebox[origin=c]{90}{\small AR}} & PP & 114.66 	{\small(1.62)} & 80.44 	{\small(.12)} &  9.27 {\small(.01)} \\
    & & GR & 108.47 {\small(1.80)} & 86.57 	{\small(.09)} & 7.87 {\small(.02)} \\
    & & PR & 119.54 {\small(1.44)} & 80.47 	{\small(.10)} &   9.27 {\small(.01)} \\
    & & KD & 148.84 {\small(2.14)} & 86.30 	{\small(.14)} &   10.41 {\small(.02)} \\
    & & PLA & 265.51 	{\small(2.53)} & 79.71 	{\small(.12)} &  14.30 {\small(.04)} \\
    & & KMD & 106.11 	{\small(1.31)} & 79.83 	{\small(.09)} &  8.93 {\small(.02)} \\
    \cline{2-6}
    & \multirow{2}{*}{ \rotatebox[origin=c]{90}{\small OS}} & AON & 126.61 & 76.40 &   9.70\\
    & & ITA & 125.44  & 76.40 &  9.71 \\
    \cline{2-6}
    \rowcolor{lavender}
    \multicolumn{1}{c}{} & \multicolumn{1}{c}{} & \bf METIS & \textbf{87.42} & \textbf{86.68} & \textbf{7.41} \\
    \midrule
    
    \multirow{9}{*}{\rotatebox[origin=c]{90}{Rome}} & \multirow{6}{*}{ \rotatebox[origin=c]{90}{\small AR}} & PP & 143.85 {\small(4.14)} & 34.44 	{\small(.07)} &    { 6.82 \small(.02)} \\
    & & GR & 138.36 	{\small(3.77)} & \textbf{51.57 	{\small(.26)}} & \textbf{ 5.41 {\small(.02)}} \\
    & & PR & 141.74 	{\small(2.70)} & 42.02 	{\small(.14)} &  6.40 {\small(.02)} \\
    & & KD & 133.95 	{\small(4.05)} & 43.61 	{\small(.11)} &   7.39 {\small(.03)} \\
    & & PLA & 197.95 	{\small(2.11)} & 43.74 	{\small(.20)} &   10.19 {\small(.04)} \\
    & & KMD & 118.89 	{\small(3.10)} & 26.90 {\small(.03)}  &  	8.58 {\small(.01)} \\
    \cline{2-6}
    & \multirow{2}{*}{ \rotatebox[origin=c]{90}{\small OS}} & AON & 124.14 & 26.31 & 8.78 \\
    & & ITA & 123.23 & 26.36 &   8.77 \\
    \cline{2-6}
    \rowcolor{lavender}
    \multicolumn{1}{c}{} & \multicolumn{1}{c}{} & \bf METIS & 
    \textbf{64.07} & 48.51  & 5.57 \\
    \bottomrule
  \end{tabular}
  \caption{Results of METIS and the baselines on CO2 emissions, road coverage (RC), and time redundancy (five-minute window). 
  Std deviation in parentheses for non-deterministic methods. 
  In bold, the lowest value of each measure and city.}
\label{tab:results}
\end{table}

KD and PLA exhibit high road coverage and time redundancy, resulting in the highest levels of CO2 emissions across all three cities. 
This is primarily because these methods have a tendency to assign trips to considerably long routes.
Despite their simplicity, AON and ITA achieve CO2 emissions comparable to edge-weight methods (PP, PR, and GR).

\begin{figure}[htb]
    \centering
    \includegraphics[width=\columnwidth]{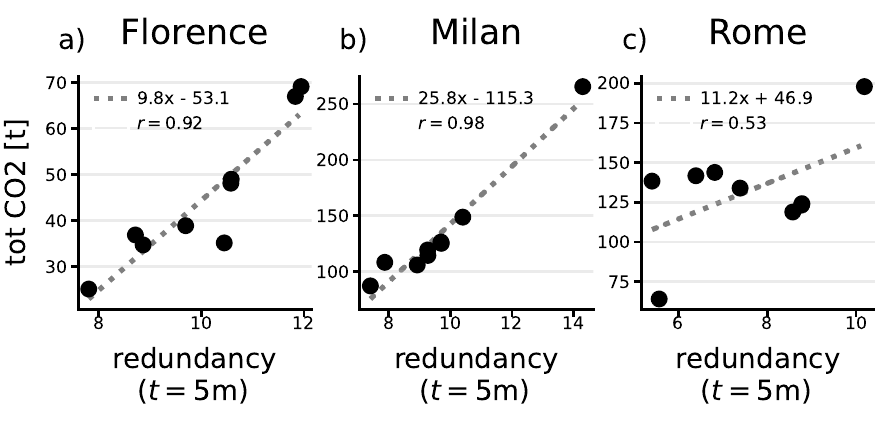}
    \vspace{-1cm}
    \caption{Pearson correlation ($r$) between time redundancy (5-minute window) and CO2 emissions. 
    Black dots represent TA algorithms. 
    The grey dashed line represents the curve fit.}
    \label{fig:corrs}
\end{figure}

\paragraph{\bf Role of time redundancy}
We find that time redundancy is crucial to assess the impact of TA solutions. 
Figure \ref{fig:corrs} shows a strong correlation between time redundancy and CO2 emissions in Florence ($r=0.92$) and Milan ($r=0.98$) and a moderate correlation in Rome ($r=0.52$). 
As the time redundancy of a TA algorithm decreases, CO2 emissions in the city also decrease: low redundancy implies that trips close in time are likely to take different routes, alleviating overlap and congestion on edges.
This means that, by utilizing the equations of Figure \ref{fig:corrs}, we can estimate the CO2 emissions of TA algorithms based solely on the characteristics of the generated routes without the need for time-consuming simulations.

\paragraph{\bf Ablation study.}
\label{sec:ablation}

To understand the role of METIS' components, we selectively remove them creating three models:

\begin{itemize}
    \item $M_1$ uses KMD and route scoring but penalizes the entire paths of vehicles in transit instead of using FLEP;
    \item $M_2$ uses KMD and route scoring but no edge penalization. 
    \item $M_3$ uses FLEP and KMD but selects among alternative routes uniformly at random. 
\end{itemize} 

\begin{figure}
    \centering
\includegraphics[width=\columnwidth]{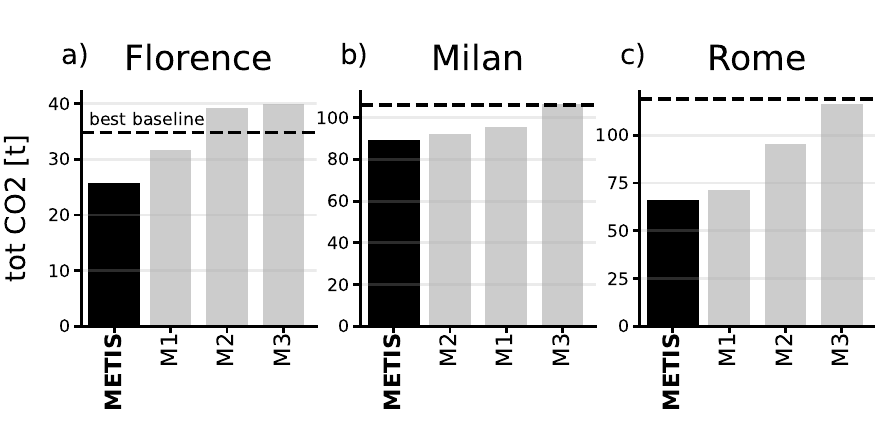}
\vspace{-1cm}
    \caption{Comparison of METIS with models based on its components ($M_1$, $M_2$, $M_3$) and the best baseline (horizontal dashed line) in terms of total CO2 emissions.}
    \label{fig:ablation}
\end{figure}

We find that removing components from METIS increases CO2 emissions compared to the complete METIS algorithm (Figure \ref{fig:ablation}). 
In Milan and Rome, $M_1, M_2, M_3$ all outperform the best baseline (KMD). 
Only $M_1$ surpasses the best baseline in Florence, while $M_2$ and $M_3$ show slightly inferior performance. 
These findings highlight the importance of the synergistic combination of METIS' components. 

\begin{figure}[htb]
    \centering
\includegraphics[width=\columnwidth]{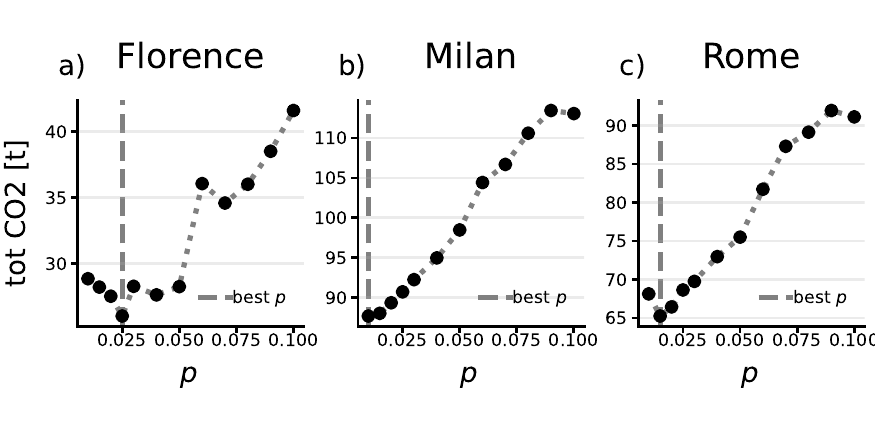}
    \vspace{-1cm}
    \caption{Relationship between METIS' parameter $p$ and CO2 emissions.
    The vertical dashed line indicates the $p$ value leading to the lowest emissions.}
    \label{fig:p_vs_co2}
    \end{figure}

\paragraph{\bf Parameter Sensitivity}
We investigate the relationship between METIS' parameter $p$, which controls penalization in FLEP, and CO2 emissions (Figure \ref{fig:p_vs_co2}). 
The analysis reveals that, apart from small values, higher values of $p$ are associated with higher CO2 emissions. 
As $p$ increases, FLEP penalizes more the edges that will be traversed by in-transit vehicles, forcing KMD to find alternative routes that may diverge considerably from the fastest path resulting in increased congestion and CO2 emissions.
In Milan and Rome, there is a clear increasing trend, showing that as $p$ increases, CO2 emissions also increase (Figure \ref{fig:p_vs_co2}b-c).
Although there is a generally increasing trend in Florence, there are multiple peaks, indicating a complex relationship between $p$ and CO2 emissions (Figure \ref{fig:p_vs_co2}a). 

We conduct a sensitivity analysis of the slowdown parameter $s$ for each city, but no significant differences were observed compared to the optimal parameter value shown in Table \ref{tab:best_params}.

\begin{figure}
    \centering
\includegraphics[width=\columnwidth]{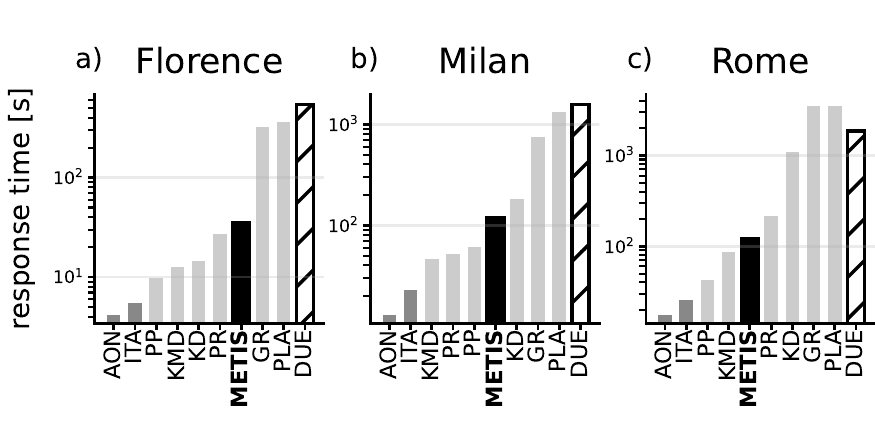}
\vspace{-1cm}
\caption{Comparison of METIS with the baselines and DUE regarding response time (seconds, in log scale) to compute TA for 1000 trips.
}
\label{fig:exetimes}
\end{figure}

\paragraph{\bf Execution times}
Figure \ref{fig:exetimes} compares METIS' response time with the baselines for 1000 trips on 16 Intel(R) Core(TM) i9-9900 CPU 3.10GHz processors with 31GB RAM on Linux 5.15.0-56-generic.

AON and ITA are the fastest approaches: the former only requires computing the fastest path; the latter involves a single weight update for each of the four splits. 
PP, PR, and KMD are the second-fastest group of baselines, while GR and PLA are the slowest. 
GR is time-consuming because it modifies the weights of every edge in the network at each iteration; PLA because it computes the shortest path trees for each trip, which is time-intensive for large graphs.
 
In general, METIS' response times are within the same order of magnitude of baselines, making it suitable for real-time TA, where both efficiency and promptness matter (Figure \ref{fig:exetimes}).

In Figure \ref{fig:exetimes}, we also show the response time of DUE (Dynamic User Equilibrium) \cite{friesz2010dynamic}, an iterative approach that approximates the user equilibrium.
DUE has considerably longer execution times than METIS when performing TA for 1000 trips: 9 minutes for Florence (14.5 times slower), 25 for Milan (11.72 times slower) and 31 minutes for Rome (14.44 times slower), see Figure \ref{fig:exetimes}.
However, this longer time does not always lead to lower CO2 emissions. 
While in Milan, DUE achieves an 18\% reduction in emissions compared to METIS, in Florence and Rome, DUE increases them by 13\% and 11\%.
These results highlight how METIS effectively reduces CO2 emissions while maintaining competitive computational performance.

\section{Conclusion}
\label{sec:conclusion}
In this paper, we introduced METIS, a one-shot cooperative algorithm for traffic assignment. 
Extensive experiments show METIS's effectiveness in reducing CO2 emissions while maintaining computational efficiency.
Future enhancements include incorporating additional measures to prioritize or discourage specific routes, refining FLEP using machine learning techniques for position estimation, estimating the slowdown factor for each road, and developing a distributed version for faster traffic assignments.

\bibliographystyle{ACM-Reference-Format}
\bibliography{biblio}




\end{document}